\documentclass[12pt]{article}

\usepackage{color}
\definecolor{darkblue}{rgb}{0.1,0.1,.7}
\usepackage[colorlinks, linkcolor=darkblue, citecolor=darkblue, urlcolor=darkblue, linktocpage]{hyperref}
\usepackage[]{amsmath}
\usepackage[]{graphicx}
\usepackage[]{latexsym}
\usepackage{slashed,graphicx,color,amsmath,amssymb}
\usepackage[mathscr]{eucal}
\usepackage{mathrsfs}
\usepackage{geometry}
\usepackage[margin=10pt,font=small,labelfont=bf]{caption}
\usepackage{amscd}
\usepackage{bm}
\usepackage{xcolor}
\usepackage{upgreek}
\usepackage[square, comma, sort&compress,numbers]{natbib}
\usepackage[all,cmtip]{xy}
\usepackage[color=cyan!30!white,linecolor=red,textsize=footnotesize]{todonotes}
\numberwithin{equation}{section}
 \makeatletter
 \g@addto@macro\bfseries{\boldmath}
 \makeatother

\usepackage{amsmath}
\begin{document}
\vspace*{-.6in} \thispagestyle{empty}
\vspace{.2in} {\Large
\begin{center}
{\bf One-loop partition functions in $T\overline{T}$-deformed AdS$_3$}
\end{center}}
\vspace{.2in}
\begin{center}
\vspace{.1in}
\renewcommand{\thefootnote}{\fnsymbol{footnote}}
\begin{center}
{Miao He\footnotemark[1]\footnotetext[1]{E-mail: hemiao@seu.edu.cn}}
\vspace{.2in}\\
\textit{School of physics \& Shing-Tung Yau Center, Southeast University\\
Nanjing 211189, P. R. China}
\end{center}
\end{center}
\vspace{.3in}

\begin{abstract}
We study the geometry of $T\bar{T}$-deformed BTZ black hole and find it can be regarded as a quotient of hyperbolic space. We then consider the massive scalar field propagating in the $T\bar{T}$-deformed BTZ black hole background. The one-loop partition function of scalar field is calculated using the heat kernel method and the Wilson spool proposal. These two methods give consistent result which implies the Wilson spool proposal still holds under $T\bar{T}$ deformation. Moreover, we also calculate the one-loop partition function of graviton in $T\bar{T}$-deformed BTZ black hole. We find the deformed one-loop partition functions are modified in a simple way, which corresponds to a replacement of the modular parameter. The result precisely matches the large $c$ expansion of $T\bar{T}$-deformed CFT partition function. These results provide a further check about the correspondence between $T\bar{T}$-deformed CFT$_2$ and AdS$_3$ with mixed boundary condition. 
\end{abstract}

\vskip 1cm \hspace{0.7cm}
\newpage
\setcounter{page}{1}
\begingroup
\hypersetup{linkcolor=black}
\tableofcontents
\endgroup
\renewcommand{\thefootnote}{\arabic{footnote}}
\section{Introduction}
\label{sec:1}
The $T\bar{T}$ deformation stems from the 2D integrable quantum field theory~\cite{Smirnov:2016lqw,Cavaglia:2016oda}. This deformation turns out to preserve the integrability and is solvable on both classical and quantum level. The deformed Lagrangian can be obtained by solving the flow equation~\cite{Bonelli:2018kik}. The deformed spectrum turns out to satisfy the inviscid Burgers equation. It is also known that $T\bar{T}$ deformation modifies the S-matrix of integrable field theory by multiplying a universal CDD factor. The exact partition function satisfies a differential equation~\cite{Datta:2018thy,Dubovsky:2018bmo}. This kind of deformation also attracts a lot of interests in string theory, 2D gravity as well as holography~\cite{Chakraborty:2018kpr,Datta:2018thy,Asrat:2017tzd,Giveon:2017nie,McGough:2016lol,Kraus:2018xrn,Cardy:2018sdv,Dubovsky:2018bmo,Callebaut:2019omt,Tolley:2019nmm,Guica:2019nzm}, see~\cite{Jiang:2019epa} for a review.
\par
On the holographic aspects, it was proposed that the $T\bar{T}$-deformed CFT$_2$ corresponds to the AdS$_3$ gravity at finite radial cutoff~\cite{McGough:2016lol}. The cutoff radial is related to the deformation parameter. In this holographic description, the $T\bar{T}$-deformed finite-size spectrum can be reproduced by calculating the quasi-local energy of the BTZ black hole. The $T\bar{T}$ flow equation coincides with the Hamilton-Jacobi equation governing the radial evolution of the classical gravity action in AdS$_3$. Another alternative holographic description is imposing a mixed boundary condition at the asymptotic AdS$_3$ boundary~\cite{Guica:2019nzm}. The AdS$_3$ solutions with mixed boundary condition can be obtained by performing a field-dependent coordinate transformation~\cite{Guica:2019nzm}. This coordinate transformation was also found from the field theory~\cite{Conti:2018tca,Conti:2019dxg}. For the mixed boundary condition holographic description, the $T\bar{T}$-deformed spectrum, Lagrangian and asymptotic asymmetries can also be reproduced in terms of Chern-Simons formulation of AdS$_3$~\cite{Ouyang:2020rpq,Llabres:2019jtx,He:2020hhm,He:2021bhj,Ebert:2022ehb}. The other holographic studies about the $T\bar{T}$ deformation can be found in~\cite{Chen:2018eqk,Caputa:2020lpa,Kraus:2021cwf,Ebert:2022cle,Kraus:2022mnu,Hirano:2020nwq,Jeong:2019ylz,Jeong:2022jmp,He:2023xnb,Chen:2023eic,Tian:2023fgf,Bhattacharyya:2023gvg,Poddar:2023ljf,Khoeini-Moghaddam:2020ymm}. Very recently, there is also a glue-on AdS holography proposal for the $T\bar{T}$-deformed CFTs~\cite{Apolo:2023vnm,Apolo:2023ckr}. 
\par
The studies about the $T\bar{T}$ deformation exhibit the new intriguing UV behaviours related to the non-locality~\cite{Guica:2021pzy}. The computations of the deformed correlation functions and entanglement entropy non-perturbatively become extremely difficult. The perturbation calculation can be found in~\cite{Guica:2019vnb,He:2019vzf,He:2020qcs,He:2020udl,He:2020cxp,He:2023kgq}. It is also shown that it is more convenient to study the correlation function in momentum space~\cite{Aharony:2018vux,Guica:2021fkv,Guica:2022gts,Chakraborty:2023wel}. The early investigation about the non-perturbative correlation function was performed by Cardy~\cite{Cardy:2018sdv}. Recent developments about the correlation functions and entanglement entropy are based on the holographic worldsheet techniques~\cite{Cui:2023jrb}, 2D gravity description of $T\bar{T}$ deformation~\cite{Aharony:2023dod} and integrability~\cite{Castro-Alvaredo:2023rtl,Castro-Alvaredo:2023wmw,Castro-Alvaredo:2023jbg,Hou:2023ytl}. The holography also provides a powerful tool to study the $T\bar{T}$-deformed CFTs from the gravity side. However, most of the holographic studies are in the large $c$ limit~\cite{Li:2020pwa,Hirano:2020ppu,He:2023hoj,He:2023knl}. In this sense, it only dual to the semi-classical limit of the AdS$_3$ gravity. We cannot even reproduce the exact $T\bar{T}$-deformed partition function from the gravity side, because the gravitational path integral is rather difficult. In addition, although we have the holographic proposals, the holographic check is only on the semi-classical level. To establish a complete holographic description of the $T\bar{T}$ deformation we have to consider the large $c$ expansion on the quantum fields theory and loop corrections of quantum gravity.
\par
In this paper, we shall step further to study the holography under $T\bar{T}$ deformation by considering the quantum gravity partition functions at one-loop correction. We use the mixed boundary condition proposal because it also holds when adding matter fields in the bulk. The gravitational path integral can be calculated by sum over all the gravitational saddles~\cite{Witten:2007kt,Yin:2007gv,Maloney:2007ud}. The semi-classical contribution is given by the Euclidean action. The one-loop correction can be obtained by considering the linear metric perturbation of the gravitational saddles. In this case, the one-loop correction can be calculated using heat kernel method at some gravitational saddles~\cite{Giombi:2008vd,David:2009xg}, see also the early studies about the scalar perturbation~\cite{Mann:1996ze,Binosi:1998yu}. Recently, there is also a Wilson spool proposal for the one-loop partition function of scalar field~\cite{Castro:2023bvo,Castro:2023dxp} in terms of Chern-Simons formulation.
\par
We start from the $T\bar{T}$-deformed BTZ black hole geometry, in which the field-dependent coordinate transformation becomes solvable. The main finding is that the $T\bar{T}$-deformed BTZ black hole is also a quotient of hyperbolic space. Since the deformed geometry is still AdS$_3$, the general techniques, such as heat kernel in AdS$_3$ and Wilson spool proposal, can be used. The heat kernel is still depends on the length of geodesics. We first consider the scalar field perturbation and find the mainly difference is the length of geodesic becomes a deformed one. The deformation of the length of geodesic leads to the fact that partition functions deformed in a simple way, which is just a replacement of the modular parameter. We then calculate the one-loop partition function of graviton. The result turns to be also a simple replacement of the modular parameter
\begin{align}
Z_{\text{graviton}}^{\text{1-loop}}=\prod_{n=2}^\infty\frac{1}{|1-q_{\lambda}^n|^2},\quad q_{\lambda}=q^{\frac{1}{\sqrt{1-\lambda\tau^2}}}\nonumber
\end{align}
which is exactly the $O(c^0)$ contribution of $T\bar{T}$-deformed CFT partition function~\cite{Datta:2021kha}. These results provide a check about correspondence between $T\bar{T}$-deformed CFT$_2$ and AdS$_3$ with mixed boundary condition at one-loop level. 
\par
The rest of the paper is organized as follows. In section~\ref{sec:2}, we show the $T\bar{T}$-deformed BTZ black hole geometry can be regarded as a quotient of hyperbolic space. In section~\ref{sec:3}, we compute one-loop partition function of scalar field in $T\bar{T}$-deformed BTZ background using both heat kernel method and Wilson spool proposal. In section~\ref{sec:4}, the one-loop graviton partition function in $T\bar{T}$-deformed BTZ background is obtained. The result turns out to be the large $c$ expansion of $T\bar{T}$-deformed CFT partition function.   The conclusion and discussion are given in section~\ref{sec:5}. In appendix~\ref{app:A}, we give a brief introduction about using the heat kernel method to calculate the path integral in AdS$_3$ . 
\section{$T\bar{T}$-deformed BTZ black hole as a quotient space}
\label{sec:2}
It was proposed that the $T\bar{T}$-deformed CFT$_2$ is dual to the AdS$_3$ gravity with mixed boundary condition~\cite{Guica:2019nzm}. The AdS$_3$ solution with mixed boundary condition can be obtained by a field-dependent coordinate transformation, we prefer to call it $T\bar{T}$-deformed AdS$_3$. In this section, we start with a brief review of the BTZ black hole as a quotient of hyperbolic space. We then show the $T\bar{T}$-deformed BTZ black hole can also be regarded as a quotient of hyperbolic space but with a different modular parameter.
\par
The most general AdS$_3$ solution with Brown-Henneaux boundary condition in Fefferman-Graham gauge~\cite{Brown:1986nw,Banados:1998gg} is called Ba\~nados geometry
\begin{align}
ds^2=\frac{dr^2}{r^2}+r^2\left(dwd\bar{w}+\frac1{r^2}\mathcal{L}dw^2+\frac1{r^2}\bar{\mathcal{L}}d\bar{w}^2+\frac1{r^4}\mathcal{L}\bar{\mathcal{L}}dwd\bar{w}\right).
\end{align}
where $\mathcal{L}$ and $\bar{\mathcal{L}}$ are holomorphic and anti-holomorphic functions, respectively. For the BTZ black hole, $\mathcal{L}$ and $\bar{\mathcal{L}}$ become a constant which related to the mass of the black hole $\mathcal{L}=\bar{\mathcal{L}}=M/2$.  The AdS$_3$ should be locally the hyperbolic space $\mathbb{H}_3$ or a quotient of $\mathbb{H}_3$ by some discrete group $\Gamma$~\cite{Kraus:2006wn}. In fact, one can perform the following coordinate transformation 
\begin{align}
y=&\frac{2 i \sqrt{\mathcal{L}} r}{r^2-\mathcal{L}}e^{-2 \sqrt{\mathcal{L}}it'},\\
\xi=&\frac{r^2+\mathcal{L}}{r^2-\mathcal{L}}e^{-2 \sqrt{\mathcal{L}}(\theta'+it')},\\
\bar{\xi}=&\frac{r^2+\mathcal{L}}{r^2-\mathcal{L}}e^{2\sqrt{\mathcal{L}}(\theta'-it')},
\end{align}
where we have used $w=\theta'+it',\bar{w}=\theta'-it'$. So that the metric becomes the Poincar\'e patch
\begin{align}
ds^2=\frac{dy^2+d\xi d\bar{\xi}}{y^2},
\end{align}
where $y>0$ and $\xi$ is a complex coordinate. We should point out that the coordinate transformation just holds for the region outside the horizon of BTZ black hole. The Poincar\'e line element can be written into the line element on $SL(2,\mathbb{C})$
\begin{align}
\label{g-metric}
ds^2=\frac{1}{2}\mathrm{Tr}(g^{-1}dgg^{-1}dg),\quad g=\left(\begin{array}{cc}y+{\xi\bar{\xi}}/{y}&{\xi}/{y}\\{\bar{\xi}}/{y}&{1}/{y}\end{array}\right)\in SL(2,\mathbb{C})
\end{align}
This line element has the following discrete group of isometry generated by $\gamma$
\begin{align}
g\to \gamma g\gamma^{\dagger},\quad \gamma=\left(\begin{array}{cc}e^{i\pi\tau}&0\\ 0 &e^{-i\pi\tau}\end{array}\right)\in SL(2,\mathbb{C}).
\end{align}
The discrete group of isometry implies the identifications of the coordinates
\begin{align}
(y,\xi)\sim(e^{i\pi(\tau-\bar{\tau})}y,e^{2i\pi\tau}\xi).
\end{align}
For the BTZ black hole, the identifications become
\begin{align}
\theta'\sim \theta+2\pi,\quad t'\sim t'+\beta_0,\quad \beta_0=\frac{\pi}{\sqrt{M}}.
\end{align}
The modular parameter is defined as the ratio of these two periods~\footnote{Here the modular parameter is a pure imaginary number, since we just consider the non-rotating BTZ black hole. If one consider the rotating BTZ black hole, there will be another period on $\theta$ direction which leads to the real part of the modular parameter.}
\begin{align}
-\frac{1}{\tau_0}=\frac{i\beta_0}{2\pi}.
\end{align}
\par
In~\cite{Guica:2019nzm}, it turns out that the $T\bar{T}$-deformed CFT corresponds to the AdS$_3$ with mixed boundary condition. The $T\bar{T}$-deformed Ba\~nados geometry can be constructed from the original Ba\~nados geometry through a field-dependent coordinate transformation, which reads
\begin{align}
dw=\frac1{1-\lambda^2\mathcal{L}_\lambda\bar{\mathcal{L}}_\lambda}(dz-\lambda\bar{\mathcal{L}}_\lambda d\bar{z}),\quad d\bar{w}=\frac1{1-\lambda^2\mathcal{L}_\lambda\bar{\mathcal{L}}_\lambda}(d\bar{z}-\lambda\mathcal{L}_\lambda dz),
\end{align}
where $\lambda$ is the deformation parameters and the $\mathcal{L}_{\lambda},\bar{\mathcal{L}}_\lambda$ are related to $\mathcal{L},\bar{\mathcal{L}}$ by
\begin{align}
\mathcal{L}=\frac{\mathcal{L}_{\lambda}(1-\lambda\bar{\mathcal{L}}_{\lambda})^2}{(1-\lambda^2\mathcal{L}_{\lambda}\bar{\mathcal{L}}_{\lambda})^2},\quad\bar{\mathcal{L}}=\frac{\bar{\mathcal{L}}_{\lambda}(1-\lambda\mathcal{L}_{\lambda})^2}{(1-\lambda^2\mathcal{L}_{\lambda}\bar{\mathcal{L}}_{\lambda})^2}.
\end{align}
The $T\bar{T}$ deformation from a point of view of field-dependent coordinate transformation was also found in~\cite{Conti:2018tca,Conti:2019dxg}. The deformed parameters $\mathcal{L}_{\lambda}$ and $\bar{\mathcal{L}}_{\lambda}$ are not the conserved charges. Following~\cite{He:2021bhj}, we find the conserved charges are related to the $T\bar{T}$-deformed spectrum and angular momentum 
\begin{align}
Q&=\frac{E_\lambda+J_\lambda}{2}=\frac{1}{2\lambda}\left[1+\lambda(\mathcal{L}-\bar{\mathcal{L}})-\sqrt{1-2\lambda(\mathcal{L}+\bar{\mathcal{L}})+\lambda^2(\mathcal{L}-\bar{\mathcal{L}})^2}\right],\\
\bar{Q}&=\frac{E_\lambda-J_\lambda}{2}=\frac{1}{2\lambda}\left[1-\lambda(\mathcal{L}-\bar{\mathcal{L}})-\sqrt{1-2\lambda(\mathcal{L}+\bar{\mathcal{L}})+\lambda^2(\mathcal{L}-\bar{\mathcal{L}})^2}\right].
\end{align}
In addition, the coordinate transformation are written in a differential form and we are not able to work out the specific form because of the coordinate dependence of two functions $\mathcal{L},\bar{\mathcal{L}}$. For the BTZ black hole, in which we can deal with the deformed metric in more details, the coordinate transformation becomes
\begin{align}
w&=\frac{(1-\lambda  Q)^2}{1-2 \lambda  Q}z-\frac{\lambda  Q (1-\lambda  Q)}{1-2 \lambda  Q}\bar{z},\\
\bar{w}&=\frac{(1-\lambda  Q)^2}{1-2 \lambda  Q}\bar{z}-\frac{\lambda  Q (1-\lambda  Q)}{1-2 \lambda  Q}z.
\end{align}
where the deformed conserved charges are
\begin{align}
Q=\bar{Q}=\frac{1-\sqrt{1-2 \lambda  M}}{2 \lambda },\quad \mathcal{L}_{\lambda}=\bar{\mathcal{L}}_{\lambda}=\frac{1-\lambda  M-\sqrt{1-2 \lambda  M}}{\lambda^2 M}.
\end{align}
After turning on the $T\bar{T}$ deformation, we can also rewrite the $T\bar{T}$-deformed BTZ black hole metric into Poincar\'e patch by employing
\begin{align}
\label{ct-1}
y=&\frac{i2r\sqrt{Q(1-\lambda Q)}}{(1-\lambda  Q)r^2 -Q}\exp\left(-\frac{2\sqrt{Q(1-\lambda Q)}}{1-2\lambda Q}it\right),\\
\label{ct-2}
\xi=&\frac{(1-\lambda  Q)r^2 +Q}{(1-\lambda  Q)r^2 -Q}\exp\left(-2\sqrt{Q(1-\lambda Q)}\theta-\frac{2\sqrt{Q(1-\lambda Q)}}{1-2\lambda Q}it\right),\\
\label{ct-3}
\bar{\xi}=&\frac{(1-\lambda  Q)r^2+Q}{(1-\lambda  Q)r^2 -Q}\exp\left(2\sqrt{Q(1-\lambda Q)}\theta-\frac{2\sqrt{Q(1-\lambda Q)}}{1-2\lambda Q}it\right).
\end{align}
where $z=\theta+it,\bar{z}=\theta-it$. The complete deformed metric can be found in other references, such as~\cite{Guica:2019nzm,He:2023xnb}. The deformed metric can also be written in the form of~\eqref{g-metric} with the $SL(2,\mathbb{C})$ group element
\begin{align}
g=\frac{i}{2r\sqrt{Q(1-\lambda Q)}}\left(\begin{array}{cc}\left(Q-r^2(1-\lambda Q)\right) e^{-\frac{2 i t \sqrt{Q(1-\lambda  Q)}}{1-2 \lambda  Q}}&\left(r^2(\lambda Q-1)-Q\right)e^{-2 \theta  \sqrt{Q (1-\lambda  Q)}} \\ \left(r^2(\lambda Q-1)-Q\right)e^{2 \theta  \sqrt{Q (1-\lambda  Q)}} &\left(Q-r^2(1-\lambda Q)\right)e^{\frac{2 i t \sqrt{Q(1-\lambda  Q)}}{1-2 \lambda  Q}}\end{array}\right).
\end{align}
Then the discrete group of isometry implies the identifications
\begin{align}
\label{dtemp}
\theta\sim \theta+2\pi,\quad t\sim t+\beta,\quad \beta=\frac{\pi(1-2\lambda Q)}{\sqrt{Q(1-\lambda Q)}}.
\end{align}
In principle, the period of the $\theta$ can be arbitrary. We assume it still has the $2\pi$ period because the $T\bar{T}$ deformation preserves the density of states and the field-dependent coordinate transformation does not change the black hole entropy~\cite{Guica:2019nzm}. The modular parameter becomes
\begin{align}
-\frac{1}{\tau}=\frac{i\beta}{2\pi}.
\end{align}
These identifications imply the $T\bar{T}$-deformed BTZ black hole can also be regarded as a quotient of hyperbolic space $\mathbb{H}_3$. The difference is that period on $t$ direction or the modular parameter becomes $\lambda$-dependent. The later of this paper is based on this result. We will consider the quantum effects of perturbation in $T\bar{T}$-deformed BTZ black hole background and compute the one-loop partition functions. 
\section{One-loop partition function of scalar field}
\label{sec:3}
In this section, we consider the massive scalar field propagating in the $T\bar{T}$-deformed BTZ black hole background. The total action becomes
\begin{align}
S=-\frac{1}{16\pi G}\int_{\mathcal{M}} d^3 x\sqrt{g}(R+2)+\frac{1}{2}\int_{\mathcal{M}} d^3 x\sqrt{g}\phi(-\Delta+m^2)\phi+B
\end{align} 
where $B$ is the boundary term depended on the boundary conditions. Then the partition function can be written in the path integral form 
\begin{align}
Z=\int[\mathcal{D}g_{\mu\nu}]e^{-I_{\mathrm{EH}}[g_{\mu\nu}]}Z_{\mathrm{scalar}}[g_{\mu\nu}],
\end{align}
where 
\begin{align}
Z_{\mathrm{scalar}}[g_{\mu\nu}]=\int[\mathcal{D}\phi]e^{-S_{\mathrm{m}}[g_{\mu\nu},\phi]}.
\end{align}
We can not be able to work out the gravitational path integral exactly. 
\par
In the semi-classical limit $G\to 0$, only the Einstein-Hilbert action and the boundary term contribute to the partition function. Under the saddle point approximation, the partition function can be calculated by sum over all the gravitational saddles. The on-shell Euclidean action becomes the energy of AdS$_3$ times the euclidean time~\cite{Maldacena:1998bw}. For the BTZ black hole, the leading semi-classical approximation to the partition function is
\begin{align}
Z_0=e^{kM\beta_0}=|q|^{-2k},\quad k=\frac{1}{4G}.
\end{align} 
For the $T\bar{T}$-deformed BTZ black hole saddles, We can obtain the similar result. In this case, the Euclidean action with boundary term would give the $T\bar{T}$-deformed spectrum~\cite{He:2020hhm}. So that the leading semi-classical approximation to the $T\bar{T}$-deformed partition function is
\begin{align}
\label{classical}
Z_0=e^{kQ\beta}=e^{\frac{k\beta }{2 \lambda }\left(1-\frac{\beta}{\sqrt{\beta ^2+4 \pi ^2 \lambda }}\right)},
\end{align}
which is agree with the result in~\cite{Apolo:2023vnm}. 
\par
The one-loop correction of partition function can be obtained by taking into account the scalar field contribution on some gravitational saddles. Usually, the BTZ black holes and thermal AdS$_3$ saddles are considered~\cite{Giombi:2008vd,David:2009xg}. The path integral for a fixed background can be calculated using the heat kernel method and the recently proposed Wilson spool method~\cite{Castro:2023bvo,Castro:2023dxp}. These considerations can also be taken for the $T\bar{T}$-deformed AdS$_3$. In what follows, we focus on the $T\bar{T}$-deformed BTZ black hole saddles. We will calculate the one-loop correction using both heat kernel method and Wilson spool method.
\subsection{Heat kernel method}
\label{sec:3.1}
The heat kernel method is a powerful technique to calculate the path integral. In terms of the heat kernel, we can write the determinant as an integral
\begin{align}
\log Z_{\mathrm{scalar}}^{\text{1-loop}}=\frac{1}{2}\int_{0}^{\infty}\frac{ds}{s}\int_{\mathcal M}d^{3}x\sqrt{g}K(s,\mathbf{x},\mathbf{x}),
\end{align}
where the heat kernel should satisfy the differential equation
\begin{align}
(\partial_s-\Delta+m^2)K(s,\mathbf{x},\mathbf{y})=0,
\end{align}
with the initial condition
\begin{align}
K(0,\mathbf{x},\mathbf{y})=\delta(\mathbf{x},\mathbf{y}).
\end{align}
For more details about the heat kernel method see~\cite{David:2009xg,Giombi:2008vd}. One can also turn to Appendix~\ref{app:A} for a brief introduction to heat kernel in AdS$_3$. The key point is to obtain the heat kernel then work out the integral to get the one-loop partition function.
\par
The information about the manifold is encoded in the Laplace operator $\Delta$ as well as the heat kernel. For the hyperbolic space $\mathbb{H}_3$, which is the maximally symmetric space, we expect the heat kernel $K^{\mathbb{H}_3}(s,\mathbf{x},\mathbf{x}')$ depends on $\mathbf{x}$ and $\mathbf{x}'$ only through the length of geodesics. Then the heat kernel for hyperbolic space $\mathbb{H}_3$ should be
\begin{align}
K^{\mathbb{H}_3}(s,\mathbf{x},\mathbf{x}')=K^{\mathbb{H}_3}(s,\sigma(\mathbf{x},\mathbf{x}'))=\frac{e^{-(m^2+1)s-\frac{\sigma^2}{4s}}}{(4\pi s)^{3/2}}\frac{\sigma}{\sinh\sigma},
\end{align}
where the length of geodesics can be calculated by
\begin{align}
\label{lgd-1}
\sigma(\mathbf{x},\mathbf{x}')&=\cosh^{-1}(1+u(\mathbf{x},\mathbf{x}')),\\
\label{lgd-2}
u(\mathbf{x},\mathbf{x}')&=\frac{(y-y')^2+|\xi-\xi'|^2}{2yy'}.
\end{align}
Now we are working in the $T\bar{T}$-deformed BTZ black hole, which turns out to be a quotient of hyperbolic space $\mathbb{H}_3/\mathbb{Z}$ in section~\ref{sec:2}.  The heat kernel for $\mathbb{H}_3/\mathbb{Z}$ can be constructed from the one for $\mathbb{H}_3$ by the method of images~\cite{David:2009xg,Giombi:2008vd}. The result turns out to be 
\begin{align}
K^{\mathrm{dBTZ}}(s,\mathbf{x},\mathbf{x}')=\sum_{n=-\infty}^{\infty}K^{\mathbb{H}_3}\left(s,\sigma(\mathbf{x},\gamma^n(\mathbf{x}'))\right),
\end{align}
where the $\gamma$ is the generators of discrete group of isometry which act on the space leads to the shift of coordinates
\begin{align}
\gamma^{n}(r,\theta,t)=(r,\theta+2\pi n,t+n\beta).
\end{align}
Therefore, the heat kernel on $\mathbb{H}_3/\mathbb{Z}$ is sum over all the geodesics winding arbitrary times around the quotient space. 
\par
For the $T\bar{T}$-deformed BTZ, in terms of $(r,\theta,t)$ coordinates, the length of geodesic $\sigma_n=\sigma(\textbf{x},\gamma^{n}(\textbf{x}))$ can be obtained by~\eqref{lgd-1}-\eqref{lgd-2} and the coordinate transformations~\eqref{ct-1}-\eqref{ct-3}. The result shows 
\begin{align}
\label{lgd}
\cosh\sigma_n=1+\left(1+\frac{r^2 (1-\lambda  Q)}{2 Q}+\frac{Q}{2r^2 (1-\lambda  Q)}\right)\sinh ^2\left(\frac{2n\pi^2}{\sqrt{\beta ^2+4 \pi ^2 \lambda}}\right),
\end{align}
where we have used $\Delta t=n \beta,\Delta\theta=2n\pi$, which means we consider the geodesics winding $n$ times around the background. We find the length of this geodesic is only related to the radial coordinate $r$. So that we can change the $r$ coordinate into the length of geodesic $\sigma_n$. Since the radial coordinate $r>0$, the range of $\sigma_n$ should be
\begin{align}
\sigma_n^0\leq \sigma_n<+\infty,\quad \sigma_n^0=\frac{4n\pi^2}{\sqrt{\beta ^2+4 \pi ^2 \lambda}}.
\end{align}
Then the integral measure of the $T\bar{T}$-deformed BTZ black hole metric becomes 
\begin{align}
d^3x\sqrt{g}=\frac{Q(1-\lambda Q)}{1-2\lambda Q}\left[\sinh ^2\left(\frac{2n\pi^2}{\sqrt{\beta ^2+4 \pi ^2 \lambda}}\right)\right]^{-1}\sinh\sigma_{n} d\sigma_{n} d\theta dt.
\end{align}
We find the integral measure only depends on the variable $\sigma_n$. Note that the $r$-dependence in the integral measure is cancelled by the Jacobian. In terms of the length of geodesics, we then can integrate the heat kernel over the spacetime.
\par
For $n=0$, the heat kernel reduce to 
\begin{align}
K^{\mathbb{H}_3}(s,\mathbf{x},\mathbf{x})=K^{\mathbb{H}_3}(s,0)=\frac{e^{-(1+m^2)s}}{(4\pi s)^{3/2}},
\end{align}
then the integral of heat kernel reads
\begin{align}
\label{n0}
\int_0^{\infty}\frac{ds}{s}\int d^3x\sqrt{g}K^{\mathbb{H}_3}(s,0)=\frac{(m^{2}+1)^{3/2}}{6\pi}\mathrm{VoL}(\text{dBTZ}).
\end{align}
The spacetime integral gives the volume of $T\bar{T}$-deformed BTZ black hole. The integral of $s$ is divergent because of the singular behaviour of the heat kernel for $s\to 0$, which can be regularized by introducing a cut-off $\epsilon$ then taking $\epsilon\to 0$. The regularized result give a prefactor in~\eqref{n0}. This divergence appears for general hyperbolic space $\mathbb{H}_3$ and $\mathbb{H}_3/\mathbb{Z}$. As argued in~\cite{Giombi:2008vd,David:2009xg}, this term can be always cancelled by a local counterterm, so that we can drop it in the later of this paper. We should note that the volume of $T\bar{T}$-deformed BTZ black hole is different from the underformed one and the deformed counterterm is needed.
\par
For $n\neq 0$, the heat kernel is given by
\begin{align}
K^{\mathbb{H}_3}(s,\sigma_n)=\frac{e^{-(1+m^2)s-\frac{\sigma_n^2}{4s}}}{(4\pi s)^{3/2}}\frac{\sigma_n}{\sinh\sigma_n}.
\end{align}
The integral of heat kernel becomes 
\begin{align}
&\int d^3x\sqrt{g}K^{\mathbb{H}_3}(s,\sigma_n)\nonumber\\
=&\int_{\sigma_{n}^0}^{\infty}d\sigma_n\int_{0}^{2\pi}d\theta\int_{0}^{\beta}dt\frac{\sigma_n e^{-(1+m^2)s-\frac{\sigma^2}{4s}}}{(4\pi s)^{3/2}}\frac{Q(1-\lambda Q)}{1-2\lambda Q}\left[\sinh ^2\left(\frac{2n\pi^2}{\sqrt{\beta ^2+4 \pi ^2 \lambda}}\right)\right]^{-1}\nonumber\\
=&\frac{e^{-(1+m^2)s-\frac{\left(\sigma_n^0\right)^2}{4s}}}{4 \pi ^{3/2} \sqrt{s}}\frac{2\pi^3}{\sqrt{\beta ^2+4 \pi ^2 \lambda}}\left[\sinh ^2\left(\frac{2n\pi^2}{\sqrt{\beta ^2+4 \pi ^2 \lambda}}\right)\right]^{-1}.
\end{align}
In the last step, we have used~\eqref{dtemp} to transform the charge $Q$ into the inverse of temperature $\beta$.
\par
Finally, the one-loop partition function of scalar field in $T\bar{T}$-deformed BTZ is 
\begin{align}
&\log Z_{\mathrm{scalar}}^{\text{1-loop}}
=\frac{1}{2}\int_0^\infty\frac{ds}s\int d^3x\sqrt{g}K^{\mathbb{H}_3}(s,0)+\frac{1}{2}\sum_{n\neq 0}\int_0^\infty\frac{ds}s\int d^3x\sqrt{g}K^{\mathbb{H}_3}(s,\sigma_n)\nonumber\\
=&\frac{(m^{2}+1)^{3/2}}{12\pi}\mathrm{VoL}(\text{dBTZ})+\sum_{n=1}^{\infty}\left[4n\sinh ^2\left(\frac{2n\pi^2}{\sqrt{\beta ^2+4 \pi ^2 \lambda}}\right)\right]^{-1}\exp\left(\frac{-4n\pi^2\sqrt{1+m^2}}{\sqrt{\beta ^2+4 \pi ^2 \lambda}}\right). 
\end{align}
The first term can be dropped as we argued before. The final result can be simplified to 
\begin{align}
Z_{\mathrm{scalar}}^{\text{1-loop}}=&\exp\left(\sum_{n=1}^{\infty}\frac{\left|q_{\lambda}\right|^{2nj}}{n\left|1-q_{\lambda}^n\right|^2}
\right)=\prod_{l,l'=0}^\infty\left(1-q_{\lambda}^{l+j}\bar{q}_{\lambda}^{l'+j}\right)^{-1},
\end{align}
where $j$ is the conformal dimension related to the mass of scalar field
\begin{align}
\label{mass-j}
j=\frac{1}{2}\left(1+\sqrt{1+m^2}\right).
\end{align}
We also have used the notation
\begin{align}
q_{\lambda}=\exp\left(\frac{4\pi^2}{\sqrt{\beta^2+4\pi^2\lambda}}\right)=q^{\frac{\beta}{\sqrt{\beta^2+4\pi^2\lambda}}},\quad q=e^{2\pi i\tau}.
\end{align}
which was first introduced in the $T\bar{T}$-deformed modular forms~\cite{Cardy:2022mhn}. This result is the same to the undeformed result in~\cite{Giombi:2008vd}, except for the replacement
\begin{align}
\beta\to\sqrt{\beta^2+4\pi^2\lambda}\quad \text{or}\quad  q\to q_{\lambda},
\end{align}
which agrees with the conclusion in~\cite{Datta:2021kha}. 
\par
The following comments about this result are in order. Firstly, the $T\bar{T}$-deformed BTZ black hole is still a quotient of hyperbolic space, which allows us to calculate the one-loop partition function using the heat kernel and method of image. The heat kernel still depends on the length of geodesics. Different from the undeformed case, the length of geodesics becomes dependent on the modular parameter in~\eqref{lgd}. Secondly, we calculate spacetime integral of the heat kernel by substituting the radial coordinate $r$ in terms of the length of geodesics $\sigma_n$. The integral range of $\sigma_n$ and spacetime integral measure are also changed.  Finally, the only differences from the undeformed case are the integral range of $\sigma_n$ and integral measure, which lead to the result deformed in a simple way. The final result shows that the $T\bar{T}$ deformation in one-loop order is just corresponds to a replacement of modular parameter. In the next section, we will see this feature is also preserved for the graviton perturbation.
\subsection{Wilson spool method}
\label{sec:3.2}
It is well-known that 3D general relativity has no local degrees of freedom, which is purely topological and can be formulated as a Chern-Simons theory~\cite{Witten:1988hc}. In the case of AdS$_3$ gravity, the relevant isometry group is $SO(2,2)\simeq SL(2,\mathbb{R})\times SL(2,\mathbb{R})$. Then Einstein-Hilbert action can be written as the difference of two copies $SL(2,\mathbb{R})$ Chern-Simons theories
\begin{align}
I_{\mathrm{EH}}[e, \omega]& = I_{CS}[A]- I_{CS}[\bar{A}],
\end{align}
where the Chern-Simons action is 
\begin{align}
I_{C S}[A]&=\frac{k}{4 \pi} \int_{\mathcal{M}} \text{Tr}\left(A \wedge d A+\frac{2}{3} A \wedge A \wedge A\right),\quad k=\frac{1}{4G}.
\end{align}
The gauge fields $A$ and $\bar{A}$ are valued in $\mathfrak{sl}(2,\mathbb{R})$, which are the linear combination of gravitational vielbein and spin connection
\begin{align}
A=\left(\omega^a+e^a\right)L_a,\quad \bar{A}=\left(\omega^a-e^a\right)L_a.
\end{align}
The $L_a$ are $\mathfrak{sl}(2,\mathbb{R})$ generators, they satisfy the commutation relations
\begin{align}
[L_a,L_b]=(a-b)L_{a+b},\quad a,b\in\{0,\pm1\}.
\end{align} 
The non-zero components of non-degenerate bilinear form are given by
\begin{align}
\operatorname{Tr}(L_0L_0)=\dfrac{1}{2},\quad\operatorname{Tr}(L_{-1}L_1)=\operatorname{Tr}(L_1L_{-1})=-1.
\end{align}
Variation of the action leads to the equations of motion  
\begin{align} 
F\equiv dA+A\wedge A=0,\quad \bar{F}\equiv d\bar{A}+\bar{A}\wedge\bar{A}=0,
\end{align}
which are equivalent to the gravitational field equation and torsion free equation. The AdS$_3$ metric can also be recovered from the gauge fields
\begin{align}
g_{ij}=\frac{1}{2}\text{Tr}\Big[(A_{i}-\bar{A}_{i})(A_{j}-\bar{A}_{j})\Big].
\end{align}
\par
Recently, it was proposed that the partition function of massive scalar field in AdS$_3$ is described by the Wilson spool in Chern-Simons formulation, which is a collection of Wilson loops winding around closed paths of the background~\cite{Castro:2023bvo,Castro:2023dxp}.
In terms of the Chern-Simons formulation, the one-loop partition function of AdS$_3$ can be obtained using the Wilson spool proposal 
\begin{align}
Z^{\text{1-loop}}_{\mathrm{scalar}}[g_{\mu\nu}]=\exp\left(\frac{1}{4}\mathbb{W}_{j}[A,\bar{A}]\right),
\end{align}
where the Wilson spool is defined as
\begin{align}
\mathbb{W}_{j}[A,\bar{A}]=i\int_{\mathcal{C}}\frac{\mathrm{d}\alpha}{\alpha}\frac{\cos\alpha/2}{\sin\alpha/2}\mathrm{Tr}_{R_j}\left(\mathcal{P}e^{\frac{\alpha}{2\pi}\oint A}\right)\mathrm{Tr}_{R_j}\left(\mathcal{P}e^{-\frac{\alpha}{2\pi}\oint \bar{A}}\right).
\end{align}
Here $j$ labels a lowest-weight representation of $\mathfrak{sl}(2,\mathbb{R})$ related to the mass of the bulk scalar field by~\eqref{mass-j}. The contour of integral for AdS$_3$ is given by $\mathcal{C}=2\mathcal{C}_{+}$ with $\mathcal{C}_{+}$ running upwards along the imaginary $\alpha$ axis to the right of zero\footnote{Since the $T\bar{T}$-deformed BTZ black hole is still AdS$_3$ solution, we choose the contour for AdS$_3$ in this paper. However, the contour of integral for dS$_3$ is given by $\mathcal{C}=\mathcal{C}_{-}\cup\mathcal{C}_{+}$, which is the union of contours running upwards along the imaginary $\alpha$ axis both to the left and right of zero. }. This contour integral becomes a sum over poles implements a sum over Wilson loops with arbitrary windings. Then the Wilson spool gives the partition function of scalar fields coupling with Chern-Simons theory. This proposal has been verified in AdS$_3$ and dS$_3$~\cite{Castro:2023bvo,Castro:2023dxp}.
\par
We now turn to the $T\bar{T}$-deformed BTZ black hole, which can be obtained from the BTZ black hole through a coordinate transformation. The AdS$_3$ gravity with mixed boundary condition can also described by two copies of $SL(2,\mathbb{R})$ Chern-Simons theory but with a non-trivial boundary term~\cite{Llabres:2019jtx,He:2020hhm}. In Chern-Simons formulation, the deformed gauge connection reads
\begin{align}
A_{r}=&\frac{1}{r}L_{0},\quad A_{\theta}=r(1-\lambda Q)L_{1}-\frac{1}{r}QL_{-1},\quad A_{t}=K\Big(r(1-\lambda Q)L_{1}-\frac{1}{r}QL_{-1}\Big),\\
-\bar{A}_{r}=&\frac{1}{r}L_{0},\quad\bar{A}_{\theta}=\frac{1}{r}QL_{1}-r(1-\lambda Q)L_{-1},\quad\bar{A}_{t}=\bar{K}\Big(\frac{1}{r}QL_{1}-r(1-\lambda Q)L_{-1}\Big), 
\end{align}
where
\begin{align}
K=-\bar{K}=\frac{1}{1-2\lambda Q}.
\end{align}
Since the $T\bar{T}$-deformed BTZ black hole geometry is filled in  $r-t$ plane, the $t$-cycle is bulk contractible while the $\theta$-cycle is not. Therefore, we choose the holonomies of the background connections $(A,\bar{A})$ around $\theta$-cycle. The Wilson loops can be calculated by diagonalizing the gauge connection
\begin{align}
\mathcal{P}\exp\left(\oint_{\gamma_{\theta}}A\right)&=u^{-1}e^{2\pi hL_0}u,\\
\mathcal{P}\exp\left(\oint_{\gamma_{\theta}}\bar{A}\right)&=\bar{u}^{-1}e^{2\pi\bar{h}L_0}\bar{u},
\end{align}
where
\begin{align}
h=-\bar{h}=&2\sqrt{Q(1-\lambda Q)}=\frac{2\pi}{\beta^2+\sqrt{4\pi^2\lambda}}.
\end{align}
So that the Wilson spool becomes 
\begin{align}
\mathbb{W}_j[a_L,a_R]=i\int_{\mathcal{C}}\frac{\mathrm{d}\alpha}{\alpha}\frac{\cos\alpha/2}{\sin\alpha/2}\chi_j\left(\frac{\alpha}{2\pi}h\right)\chi_j\left(-\frac{\alpha}{2\pi}\bar{h}\right),
\end{align}
where the characters are given by
\begin{align}
\chi_j(z)=\operatorname{Tr}_{R_j}\left(e^{2\pi zL_0}\right)=\frac{e^{\pi z(2j-1)}}{2\sinh(-\pi z)}.
\end{align}
The integral in the Wilson spools can be calculated by taking the residues of the poles. We deform the contour to the
right and pick the residues of the poles at $\alpha=2n\pi,n\in \mathbb{Z}^+$ for each $\mathcal{C}_{+}$. The result shows 
\begin{align}
\mathbb{W}_j[a_L,a_R]=&\sum_{n=1}^\infty \frac{1}{n}\chi_j(n\pi h)\chi_j(-n\pi \bar{h})\nonumber\\
=&\sum_{n=1}^{\infty}\left[n\sinh ^2\left(\frac{2n\pi^2}{\sqrt{\beta ^2+4 \pi ^2 \lambda}}\right)\right]^{-1}\exp\left(\frac{-4n\pi^2(2j-1)}{\sqrt{\beta ^2+4 \pi ^2 \lambda}}\right)\nonumber\\
=&4\log\prod_{l,l'=0}^\infty\left(1-q_{\lambda}^{l+j}\bar{q}_{\lambda}^{l'+j}\right)^{-1},
\end{align}
We then have 
\begin{align}
Z^{\text{1-loop}}_{\mathrm{scalar}}[g_{\mu\nu}]=\exp\left(\frac{1}{4}\mathbb{W}_{j}[A,\bar{A}]\right)=
\prod_{l,l'=0}^\infty\left(1-q_{\lambda}^{l+j}\bar{q}_{\lambda}^{l'+j}\right)^{-1}.
\end{align}
This is the same result we have obtained using the heat kernel method. Therefore, the Wilson spool proposal still holds for the $T\bar{T}$-deformed AdS$_3$. 
\par
From the calculation we find the Wilson spool is more convenient since we do not need to treat more details about the length of geodesics and the periodicity of $\theta$ and $t$ directions. These information is encoded in the Wilson loops winding around the background. The method of image is becomes sum over all the Wilson loops which exactly corresponds to the Wilson spool proposal. Moreover, the Wilson spool also gives the character expansion of the one-loop partition function.
\section{One-loop partition function of graviton}
\label{sec:4}
We now consider the graviton perturbation of the $T\bar{T}$-deformed BTZ black hole. Since the $T\bar{T}$-deformed AdS$_3$ metric still satisfy the Einstein equation with a negative cosmological constant, we can use the same technique in AdS$_3$. We should consider the perturbation
\begin{align}
g_{\mu\nu}\to g_{\mu\nu}+h_{\mu\nu},
\end{align}
and find the action for the metric $h_{\mu\nu}$. We assume the perturbation $h_{\mu\nu}$ is the order of $O(\sqrt{G})$, so it would contribute the one-loop correction. The kinetic term for $h_{\mu\nu}$ depends on the choice of gauge. Under the gauge of~\cite{Christensen:1979iy}, the result shows the perturbative action can be separated into trace part and traceless part, which is described by the scalar field and the traceless symmetric tensor field, defined as
\begin{align}
\phi=h_{\rho}^{\rho},\quad \phi_{\mu\nu}=h_{\mu\nu}-\frac{1}{3}g_{\mu\nu}h^{\rho}_{\rho}.
\end{align}
In addition, the gauge-fixing procedure also introduces a Fadeev-Popov ghost field, which in this case is a complex valued vector field $\phi_{\mu}$. Including the graviton perturbation, we therefore have the action
\begin{align}
S=I_{\mathrm{EH}}[g_{\mu\nu}]+S_{\mathrm{trace}}[\phi]+S_{\mathrm{traceless}}[\phi_{\mu\nu}]+S_{\mathrm{ghost}}[\phi_{\mu}],
\end{align}
where
\begin{align}
S_{\mathrm{trace}}&=-\frac{1}{32\pi G}\int d^3x\sqrt{g}\left[\frac{1}{12}\phi\left(-\Delta+4\right)\phi\right],\nonumber\\
S_{\mathrm{traceless}}&=-\frac{1}{32\pi G}\int d^3x\sqrt{g}\left[\frac{1}{2}\phi_{\mu\nu}\left(g^{\mu\rho}g^{\nu\sigma}\Delta+2R^{\mu\rho\nu\sigma}\right)\phi_{\rho\sigma}
\right],\nonumber\\
S_{\mathrm{ghost}}&=\frac{1}{32\pi G}\int d^3x\sqrt{g}\phi^{*}_{\mu}\left(-g^{\mu\nu}\Delta-R^{\mu\nu}\right)\phi_{\nu}.
\end{align}
\par
Then the one-loop partition function contains three parts
\begin{align}
\log Z_{\text{graviton}}^{\text{1-loop}}&=- \frac{1}{2}\log\det\Delta^{(2)}+\log\det\Delta^{(1)}-\frac{1}{2}\log\det\Delta^{(0)},
\end{align}
where
\begin{align}
\Delta^{(0)}=&-\Delta+4,\\
\Delta^{(1)}=&(\Delta+2)\delta_{\mu}^{\nu},\\
\Delta^{(2)}=&\left(\Delta+2\right)\delta_\mu^\rho\delta_\nu^\sigma.
\end{align}
The determinate can also be calculated by the heat kernel method in quotient space
\begin{align}
-\log\det\Delta^{(0)}&=\sum_{n=-\infty}^{\infty}\int_0^{\infty}\frac{ds}{s}\int d^3x\sqrt{g}K^{\mathbb{H}_3}(s,\sigma_n),\\
-\log\det\Delta^{(1)}&=\sum_{n=-\infty}^{\infty}\int_0^{\infty}\frac{ds}{s}\int d^3x\sqrt{g}\hat{g}^{\mu\nu^{\prime}}K_{\mu\nu^{\prime}}^{\mathbb{H}_3}(s,\sigma_n),\\
-\log\det\Delta^{(2)}&=\sum_{n=-\infty}^{\infty}\int_0^{\infty}\frac{ds}{s}\int d^3x\sqrt{g}\hat{g}^{\mu\mu'}\hat{g}^{\nu\nu'}K_{\mu\nu,\mu'\nu'}^{\mathbb{H}_3}(s,\sigma_n),
\end{align}
where the metric is defined as 
\begin{align}
\hat{g}^{\mu\nu^{\prime}}\equiv g^{\mu\rho}(x)\frac{\partial(\gamma^{n}x)^{\nu^{\prime}}}{\partial x^{\rho}}.
\end{align}
In our case, the action of $\gamma$ is just a translation of the coordinates, so the $\hat{g}$ is the same with $g$. The heat kernel for each operators should satisfy
\begin{align}
\partial_{s}K_{\mu\nu'}(s,\mathbf{x},\mathbf{x}')+\left(\Delta^{(1)}\right)_{\mu}^{\nu}K_{\nu\nu'}(s,\mathbf{x},\mathbf{x}')=0,\\
\partial_sK_{\mu\nu,\mu'\nu'}(s,\mathbf{x},\mathbf{x}')+\left(\Delta^{(2)}\right)_{\mu\nu}^{\rho\sigma}K_{\rho\sigma,\mu'\nu'}(s,\mathbf{x},\mathbf{x}')=0.
\end{align}
While the operator $\Delta^{(0)}$ is the same as scalar field case with the mass $m^2=4$. For the hyperbolic space, the heat kernel only depends on the length of geodesics, so that the heat kernel can be obtained exactly. The expressions for the heat kernels are rather complicated, which can be found in~\cite{Giombi:2008vd,David:2009xg}. 
\par
Note that one of the important features is that the determinant only depends on the trace of heat kernel, which is diffeomorphism invariant. Therefore, the trace of the heat kernel for $T\bar{T}$-deformed BTZ black hole is the same as the one for BTZ black hole except for the length of geodesics $\sigma_n$.  Similar to the scalar field case, the length of geodesics becomes a deformed one, which would lead to a change of spacetime integral measure and the range of $\sigma_n$. After taking into account the change of geodesics, we can obtain the integral of the trace of heat kernel over the spacetime. The one-loop determinant for the traceless symmetric tensor field turns to be  
\begin{align}
&\int d^3x\sqrt{g}\hat{g}^{\mu\mu'}\hat{g}^{\nu\nu'}K_{\mu\nu,\mu'\nu'}^{\mathbb{H}_3}(s,\sigma_n)\nonumber\\
=&\frac{e^{-\frac{\left(\sigma_n^0\right)^2}{4s}}}{2\pi^{\frac32}\sqrt{s}}\left(e^{-s}+e^{-4s}+\frac{e^{-5s}}{2}\right)\frac{2\pi^3}{\sqrt{\beta ^2+4 \pi ^2 \lambda}}\left[\sinh ^2\left(\frac{2n\pi^2}{\sqrt{\beta ^2+4 \pi ^2 \lambda}}\right)\right]^{-1}.
\end{align}
The contribution from the vector ghost can be obtained analogously 
\begin{align}
&\int d^3x\sqrt{g}\hat{g}^{\mu\nu^{\prime}}K_{\mu\nu^{\prime}}^{\mathbb{H}_3}(s,\sigma_n)\nonumber\\
=&-\frac{e^{-\frac{(\sigma_n^0)^2}{4s}}}{2\pi^{\frac32}\sqrt{s}}\left(2e^{-4s}+e^{-5s}\right)\frac{2\pi^3}{\sqrt{\beta ^2+4 \pi ^2 \lambda}}\left[\sinh ^2\left(\frac{2n\pi^2}{\sqrt{\beta ^2+4 \pi ^2 \lambda}}\right)\right]^{-1}.
\end{align}
The contribution from the traceless part is just the scalar field case with mass $m^2=4$ 
\begin{align}
&\int d^3x\sqrt{g}K^{\mathbb{H}_3}(s,\sigma_n)
=\frac{e^{-\frac{\left(\sigma_n^0\right)^2}{4s}-5s}}{4\pi^{\frac32}\sqrt{s}}\frac{2\pi^3}{\sqrt{\beta ^2+4 \pi ^2 \lambda}}\left[\sinh ^2\left(\frac{2n\pi^2}{\sqrt{\beta ^2+4 \pi ^2 \lambda}}\right)\right]^{-1}.
\end{align}
We also have to integrate the parameter $s$. Similarly to the scalar field case, the integral of $s$ is divergent for $n=0$ and can be always cancelled by a conuterterm. The effective contribution comes from the $n\neq 0$ case.
\par
The final result turns out to be
\begin{align}
\log Z_{\text{graviton}}^{\text{1-loop}}
&=\sum_{n=1}^{\infty}\int_0^{\infty}\frac{ds}{s}\frac{e^{-\frac{\left(\sigma_n^0\right)^2}{4s}}}{4\pi^{\frac32}\sqrt{s}}\left(e^{-s}-e^{-4s}\right)\frac{2\pi^3}{\sqrt{\beta ^2+4 \pi ^2 \lambda}}\left[\sinh ^2\left(\frac{2n\pi^2}{\sqrt{\beta ^2+4 \pi ^2 \lambda}}\right)\right]^{-1}\nonumber\\
&=\sum_{n=1}^\infty\frac{2}{n}\left(\frac{\left|q_{\lambda}\right|^{2n}}{\left|1-q_{\lambda}^n\right|^2}-\frac{\left|q_{\lambda}\right|^{3n}}{\left|1-q_{\lambda}^n\right|^2}\right) \\
&=-\sum_{n=2}^\infty\log|1-q_{\lambda}^n|^2
\end{align}
Therefore we have 
\begin{align}
Z_{\text{graviton}}^{\text{1-loop}}=\prod_{n=2}^\infty\frac{1}{|1-q_{\lambda}^n|^2},
\end{align}
which precisely match the result in~\cite{Datta:2021kha}. In~\cite{Datta:2021kha}, the authors consider large $c$ expansion of the $T\bar{T}$-deformed CFT partition function by taking the double scaling limit
\begin{align}
c\to\infty,\quad\lambda\to0,\quad c\lambda=\mathrm{fixed}.
\end{align} 
The $O(c^0)$ contribution strikingly turns out to be of the same form as the undeformed case except for the change of modular parameter.
\par
The full partition function is therefore given by multiplying the tree level contribution~\eqref{classical}, which reads
\begin{align}
Z_{\text{gravity}}=|q_{\lambda}|^{-2k_{\lambda}}\prod_{n=2}^\infty\frac{1}{|1-q_{\lambda}^n|^2},\quad k_{\lambda}=\frac{\beta  k}{2 \pi ^2 \lambda }\left(\sqrt{\beta ^2+4 \pi ^2 \lambda }-\beta \right).
\end{align}
This result has the same formula with the undeformed case, except for the replacements $k\to k_{\lambda}$ and $q\to q_{\lambda}$. Similar to the undeformed case, it can be interpreted as 
\begin{align}
Z_{\text{gravity}}=\text{Tr}q_{\lambda}^{L_{0}}\bar{q}_{\lambda}^{\bar{L}_{0}},
\end{align}
where the representation of Virasoro algebra contains a ground state $L_0|0\rangle=-k_{\lambda}|0\rangle $ and its Virasoro descendants~\cite{Maloney:2007ud}. For the undeformed case, the result comes from the  observation of that the asymptotic symmetry group of AdS$_3$ with Brown-Henneaux boundary condition is the Virasoro algebra~\cite{Brown:1986nw}. While the $T\bar{T}$-deformed case can be interpreted as that the asymptotic symmetry group associated with the mixed boundary condition is a state-dependent Virasoro algebra~\cite{Guica:2019nzm,He:2021bhj}. 
\section{Conclusion and discussion}
\label{sec:5}
In this work, we consider the AdS$_3$ with mixed boundary condition, whose solution can be obtained by a field-dependent coordinate transformation. We found the $T\bar{T}$-deformed BTZ black hole can be regarded as a quotient of hyperbolic space. Based on this fact, we studied the scalar field propagating in $T\bar{T}$-deformed BTZ black hole background. The one-loop partition function for scalar field in $T\bar{T}$-deformed BTZ black hole was calculated using the heat kernel method and Wilson spool method. We also computed the one-loop partition function for graviton by considering the tensor perturbation of $T\bar{T}$-deformed BTZ black hole. We found the one-loop partition function is deformed in a simple way, which is just a replacement of the modular parameter. These result coincide with the $T\bar{T}$-deformed CFT partition function in a special double scaling limit of large central charge and small deformation parameter. Including the tree level contribution, the full $T\bar{T}$-deformed gravity partition function also have a similar form with the undeformed one, which agree with the asymptotic symmetry group associated with the mixed boundary condition is a state-dependent Virasoro algebra. 
\par
For the scalar one-loop partition function, we found the Wilson spool proposal is more convenient. However, for the gravity one-loop partition function, we only used the heat kernel method since we do not have the Wilson spool proposal for the $U(1)$ gauge field and traceless symmetric tensor field. From the heat kernel calculations we found the Wilson spool proposal~\cite{Castro:2023bvo} may be generalized to the gauge field by considering an appropriate representation of $SL(2,\mathbb{R})$. It would be interesting to generalize the Wilson spool proposal to the higher spin fields in AdS$_3$. 
\par
It is worth noting that the one-loop partition function is not modular invariant since we just consider the $T\bar{T}$-deformed BTZ black hole saddle. To obtain the full modular invariant partition function we have to sum over all the saddles including BTZ black hole, thermal AdS and BTZ black hole with conical singularity. In present, we do not know how to obtain the deformed thermal AdS and the deformed BTZ black hole with conical singularity. For the undeformed case, the thermal AdS partition function can be obtained by the modular transformation. For the $T\bar{T}$-deformed case, the modular invariant turns to be related to the transformation of the deformation parameter~\cite{Aharony:2018bad,Aharony:2018ics}. Using the modular transformation with associated transformation of deformation parameter, we can obtain contributions from other saddles.
\par
The one-loop partition is also related to the correlation functions in the context of AdS$_3$/CFT$_2$~\cite{Chen:2015uga}. The one-loop partition function can be written into a summation of correlation functions. The one-loop partition function formula provides us with a result to infer the two-point function and higher point function in $T\bar{T}$-deformed CFTs.
\section*{Acknowledgements}
I am grateful to Yunfeng Jiang for helpful discussions and comments on this manuscript. I also thank Yuan Sun for helpful discussions. MH is supported by China Postdoctoral Science Foundation No.2023M740612. It is a pleasure to thank the Yukawa Institute for Theoretical Physics for hospitality in Kyoto during the 18th Kavli Asian Winter School on Strings, Particles and Cosmology (YITP-W-23-13), where part of this work was done.
\appendix
\section{Heat kernel method}
\label{app:A}
The heat kernel method is used to calculate the path integral. Here we give a brief introduction about this method. We start from the massive free field theory, whose action is given by 
\begin{align}
S(\phi)=\frac{1}{2}\int_{\mathcal{M}}d^3x{\sqrt{g}}\phi(\Delta+m^2)\phi.
\end{align}
Then the partition function can be written as a path integral form
\begin{align}
Z=\int [\mathcal{D}\phi] e^{-S(\phi)}.
\end{align}
This is a Gaussian integral, which can be written as
\begin{align}
\log Z=-\frac12\log\det(\Delta+m^2)=-\frac12\sum_n\log\lambda_n,
\end{align} 
where the $\lambda_n$ is the eigenvalue of operator. The most straightforward procedure is to find a complete basis of orthogonal normalized eigenfunctions 
\begin{align}
\sum_n\psi_n(\mathbf{x})\psi_n(\mathbf{x})=\delta^3(\mathbf{x},\mathbf{y}),\quad&\int_{\mathcal{M}}d^3x\sqrt{g}\psi_n(\mathbf{x})\psi_m(\mathbf{x})=\delta_{nm},
\end{align}
which obeys $(\Delta+m^2)\psi_n=\lambda_n\psi_n$.
The heat kernel is constructed by
\begin{align}
K(s,\mathbf{x},\mathbf{y})=\sum_ne^{-\lambda_n s}\psi_n(\mathbf{x})\psi_n(\mathbf{y}).
\end{align}
One can verify the heat kernel satisfies the differential equation
\begin{align}
\label{hk-equation-a}
(\partial_s+\Delta_{\mathbf{x}}+m^2)K(s,\mathbf{x},\mathbf{y})=0,
\end{align}
with the initial condition
\begin{align}
K(0,\mathbf{x},\mathbf{y})=\delta(\mathbf{x},\mathbf{y}).
\end{align}
In practice, we may obtain the heat kernel as the unique solution to the differential equation with initial condition. In terms of the heat kernel, the sum over the eigenvalues of Laplace operator can be written as a integral
\begin{align}
\log Z_{\mathrm{scalar}}=\frac{1}{2}\int_{0}^{\infty}\frac{ds}{s}\int_{\mathcal M}d^{3}x\sqrt{g}K(s,\mathbf{x},\mathbf{x}).
\end{align}
This method can also be used for the spinors, vector fields and symmetric traceless tensor fields, etc.
\par
For the $U(1)$ gauge field, the heat kernel is a bi-tensor $K_{\mu\nu'}(s,\mathbf{x},\mathbf{x}')$, which satisfies the differential equation
\begin{align*}
\partial_{s}K_{\mu\nu'}(s,\mathbf{x},\mathbf{x}')+(\Delta^{(1)})_{\mu}^{\nu}K_{\nu\nu'}(s,\mathbf{x},\mathbf{x}')=0
\end{align*}
with the initial condition
\begin{align}
K_{\mu\nu'}(0,\mathbf{x},\mathbf{x}')=g_{\mu\nu'}(\mathbf{x})\delta^3(\mathbf{x},\mathbf{x}').
\end{align}
The partition function for gauge field can be calculated as 
\begin{align}
\log Z_{\mathrm{gauge}}=\frac{1}{2}\int_{0}^{\infty}\frac{ds}{s}\int_{\mathcal M}d^{3}x\sqrt{g}g^{\mu\nu'}K_{\mu\nu'}(s,\mathbf{x},\mathbf{x}').
\end{align}
\par
For the traceless symmetric tensor field, the heat kernel is $K_{\mu\nu,\mu'\nu'}(s,\mathbf{x},\mathbf{x}')$, which satisfies the differential equation
\begin{align*}
\partial_{s}K_{\mu\nu,\mu'\nu'}(s,\mathbf{x},\mathbf{x}')+(\Delta^{(2)})_{\mu\nu}^{\rho\sigma}K_{\rho\sigma,\mu'\nu'}(t,\mathbf{x},\mathbf{x}')=0
\end{align*}
with the initial condition
\begin{align}
K_{\mu\nu,\mu'\nu'}(0,\mathbf{x},\mathbf{x}')=\frac{1}{2}\left(g_{\mu\mu'}g_{\nu\nu'}+g_{\mu\nu'}g_{\nu\mu'}-\frac{2}{3}g_{\mu\nu}g_{\mu'\nu'}\right)\delta^3(\mathbf{x},\mathbf{x}').
\end{align}
Since the field is traceless, the heat kernel should be traceless as well
\begin{align}
g^{\mu\nu}K_{\mu\nu,\mu'\nu'}(s,\mathbf{x},\mathbf{x}')=g^{\mu'\nu'}K_{\mu\nu,\mu'\nu'}(s,\mathbf{x},\mathbf{x}')=0
\end{align}
The partition function for gauge field can be calculated as 
\begin{align}
\log Z_{\mathrm{traceless}}=\frac{1}{2}\int_{0}^{\infty}\frac{ds}{s}\int_{\mathcal M}d^{3}x\sqrt{g}g^{\mu\mu'}g^{\nu\nu'}K_{\mu\nu,\mu'\nu'}(s,\mathbf{x},\mathbf{x}')
\end{align}
In this method, the key point is to find the heat kernel in various manifold.  We then show two examples of the manifold $\mathcal{M}$, which is helpful in this paper.
\paragraph{Hyperbolic Space}
We use the following metric on Hyperbolic Space $\mathbb{H}_3$
\begin{align}
ds^2=\frac{dy^2+d\xi d\bar{\xi}}{y^2},
\end{align}
where $y>0$ and $\xi$ is a complex coordinate. Since the hyperbolic space is maximally symmetric, we expect the heat kernel $K(s,\mathbf{x},\mathbf{x}')$ just depends on $\mathbf{x}$ and $\mathbf{x}'$ only through the length of geodesics
\begin{align}
\sigma(\mathbf{x},\mathbf{x}')&=\cosh^{-1}(1+u(\mathbf{x},\mathbf{x}')),\\
u(\mathbf{x},\mathbf{x}')&=\frac{(y-y')^2+|\xi-\xi'|^2}{2yy'}.
\end{align}
In terms of the length of geodesic, the Laplace operator becomes 
\begin{align}
\Delta=\partial_{\sigma}^2+2\coth\sigma\partial_\sigma.
\end{align}
In this case, it is straightforward to solve the differential equation~\eqref{hk-equation-a} and obtain the heat kernel
\begin{align}
K^{\mathbb{H}_3}(s,\sigma)=\frac{e^{-(m^2+1)s-\frac{\sigma^2}{4s}}}{(4\pi s)^{3/2}}\frac{\sigma}{\sinh\sigma}.
\end{align}
\par
The heat kernel for the $U(1)$ gauge field and the traceless symmetric tensor field in $\mathbb{H}_3$ also depends on the length of geodesics. The complete expression for the heat kernel is rather complicated, we prefer to write down the trace of these two heat kernels. For the $U(1)$ gauge field, we have
\begin{align}
g^{\mu\nu^{\prime}}K_{\mu\nu^{\prime}}^{\mathbb{H}_{3}}(s,\sigma) =-3\left(F(t,\sigma)+\frac{1}{\sinh\sigma}\partial_{\sigma}S(t,\sigma)\right)
\end{align} 
where
\begin{align}
F(s,\sigma)&=-\frac{e^{-\frac{\sigma^2}{4s}}}{(4\pi s)^{3/2}}\frac{\sigma}{\sinh\sigma},\\
S(s,\sigma)&=\frac{4}{(4\pi)^{3/2}}\frac{e^{-\frac{\sigma^2}{4s}}}{\sinh\sigma}\sqrt{s}\int_0^1d\xi e^{-s(1-\xi)^2}\sinh\sigma\xi.
\end{align}
For the traceless symmetric tensor field, the heat kernel becomes more complicated, which can be found in the literatures~\cite{Giombi:2008vd,David:2009xg}. We do not write down the concrete expression here.
\paragraph{BTZ black hole}
The BTZ black hole is described by the quotient space geometry of hyperbolic space $\mathbb{H}_3/\mathbb{Z}$. The quotient space can be obtained by the identification 
\begin{align}
\gamma_0^{n}(\rho,\theta,t)\to(\rho,\theta+2\pi n,t+n\beta_0).
\end{align}
where the $\gamma_0$ is the generators of discrete group of isometry. We also expect the heat kernel $K(s,\mathbf{x},\mathbf{x}')$ just depends on the length of geodesics between $\mathbf{x}$ and $\mathbf{x}'$. The heat kernel on $\mathbb{H}_3/\mathbb{Z}$ can be obtained from the one on $\mathbb{H}_3$ by the method of images
\begin{align}
K^{\mathbb{H}_3/\mathbb{Z}}(s,\mathbf{x},\mathbf{x}')=\sum_{n=-\infty}^{\infty}K^{\mathbb{H}_3}\left(s,\sigma_n\right),\quad  \sigma_n\equiv\sigma(\mathbf{x},\gamma_0^n(\mathbf{x}'))
\end{align}
Therefore, the heat kernel on $\mathbb{H}_3/\mathbb{Z}$ is sum over all the geodesics winding arbitrary times around the quotient space. This method can also be used for the heat kernel for gauge field and traceless symmetric tensor field.
\bibliographystyle{utphys}
\bibliography{reference}
\end{document}